\documentclass[twocolumn,showpacs,preprintnumbers,amsmath,amssymb]{revtex4}
\usepackage{dcolumn}
\usepackage{bm}
\usepackage[dvips]{graphicx}

\def\etal{{\it et~al.,\ }}
\def\RB{I\hspace{-0.5mm}I}
\def\RC{I\hspace{-0.5mm}I\hspace{-0.5mm}I}
\def\RD{I\hspace{-0.5mm}V}
\def\etal{{\it et~al.,\ }}
\begin{document}
\title{Phase Transitions in the Bilayer $\nu =2/3$ Quantum Hall Effect}
\author{N. Kumada}
\email{kumada@mail.cc.tohoku.ac.jp}
\author{D. Terasawa}
\author{Y. Shimoda}
\author{H. Azuhata}
\author{A. Sawada}
\author{Z. F. Ezawa}
\affiliation{Department of Physics, Tohoku University, Sendai 980-8578, Japan}
\author{K. Muraki}
\author{T. Saku}
\author{Y. Hirayama}
\affiliation{NTT Basic Research Laboratories, 3-1 Morinosato-Wakamiya, Atsugi, Kanagawa 243-0198, Japan}

\date{Version: \today}

\begin{abstract}
We measured the magnetoresistance of bilayer quantum Hall (QH) effects at the fractional filling factor $\nu =2/3$ by changing the total electron density and the density difference between two layers.
Three different QH states were separated by two types of phase transition: One is the spin transition and the other is the pseudospin transition.
In addition, two different hystereses were detected, one of which is specific to bilayer systems.
The phase transitions and the hystereses are described well by a composite fermion model extended to a bilayer system.
\end{abstract}
\maketitle


In bilayer electron systems, interlayer Coulomb and tunnelling interactions provide an additional degree of freedom, which produce rich phenomena with no counterpart in individual two dimensional systems \cite{spielman}.
A good example is the bilayer quantum Hall (QH) state at the filling factor $\nu =2$.
A phase transition has been observed at $\nu =2$ as revealed by magnetotransport measurements \cite{SawadaPRL,SawadaPRB}, light-scattering \cite{Pellegrini,Pellegriniscience} and capacitance spectroscopy \cite{Khrapai}.
In the weak interlayer correlation case, electrons in each layer configure the monolayer $\nu =1$ QH state separately.
This state is the compound state with $\nu =1+1$, which is a spin-polarized state.
When the interlayer correlation is enhanced, this compound state transits to a spin-unpolarized QH state.
In a conventional system, cyclotron energy is very large \cite{cyc} and the Landau orbital degree of freedom is frozen.
Thus, the observed phase transition is due to the spin and layer degree of freedom.

The fractional QH effect (FQHE) is intuitively understood based on the composite fermion (CF) model \cite{Jain}, where basic particles are CFs obtained by attaching an even number of flux quanta to electron.
At the filling factor $\nu =1/2$, the attached flux exactly cancels the applied magnetic field.
As the field deviates from $\nu =1/2$, CFs are subjected to the effective magnetic field $B^\ast $ and quantized into CF Landau levels.
The series of FQHE at $\nu =p/(2p\pm 1)$ are interpreted as integer QH effect (IQHE) of CF at $\nu _{CF}=p$.
For instance, the $\nu =2/3$ FQHE is mapped to the $\nu =2$ IQHE.
However, there exists significant difference because the CF-cyclotron gap is comparable to the Zeeman energy.
In the monolayer QH states, indeed, phase transitions have been observed \cite{Du}, which arise from a competition between the Zeeman and CF-cyclotron energies.
Moreover, the hystereses \cite{Kronmuller1,Kronmuller2,Cho,Eom,Smet,Hashimoto} observed around $\nu =2/3$ and $\nu =2/5$ in monolayer systems have attracted much attention recently.
Thus we expect rich phases to appear in the bilayer $\nu =2/3$ QH state, where CF-Landau orbital, spin and layer degrees of freedom all come into play.

In this Letter, we study the magnetoresistance at $\nu =2/3$ in a bilayer system by changing the total electron density $n_t$ and the density difference between two layers $\sigma =(n_f-n_b)/n_t$ and map out the phase diagram in the $n_t$-$\sigma $ space, where $n_t$ ($n_b$) is the electron density in the front (back) layer.
In the phase diagram, three different QH states and a no-QH area were observed.
The QH states were elucidated by the activation energy measurements.
In addition, two types of hysteresis were detected.
One of the hystereses, specific to bilayer systems, is our new finding.
These phases are described well by a CF model extended to a bilayer system on the understanding that collective Coulomb interactions renormalize not only the CF cyclotron energy but also the Zeeman and tunnelling energies.

Our sample was grown by molecular beam epitaxy on a (100)-oriented GaAs substrate.
It consists of two GaAs quantum wells of 200\,\AA\ width separated by a 31-\AA-thick barrier of Al$_{0.33}$Ga$_{0.67}$As \cite{Muraki}.
The tunnelling energy gap is $\Delta _{SAS}=10.9$\,K.
The electron density in each layer is controlled by adjusting the front and back gate voltages.
The low temperature mobility is 2$\times 10^{6}$\,cm$^2/$Vs with $n_t=2\times 10^{11}$\,cm$^{-2}$.
Measurements were performed with the sample mounted on a goniometer with a superconducting stepper motor \cite{GONO} in a dilution refrigerator.
Standard low-frequency ac lock-in techniques were used.
All magnetoresistance data were taken at 50\,mK with the magnetic field sweep rate 0.06\,T/min and the current 20\,nA.

\begin{figure}[t]
\begin{center}
\includegraphics[width=70mm]{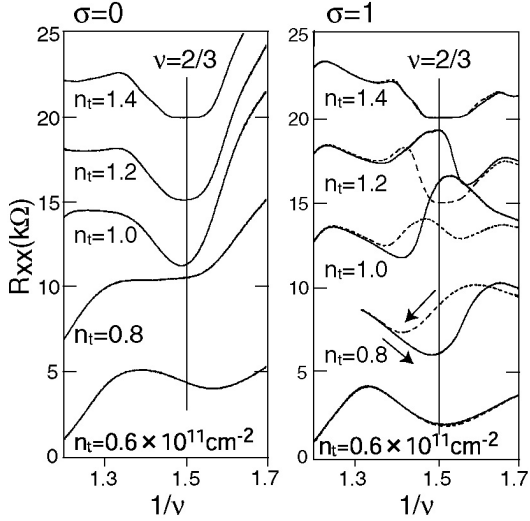}
\caption{\small
Magnetoresistances at the balanced density point $\sigma =0$ and at the monolayer point $\sigma =1$ for several total densities.
They are plotted as a function of $1/\nu =eB_\perp /hn_t$.
Traces are vertically offset by 5\,k$\Omega $ for each curve.
Hysteresis between the upward (solid trace) and downward (dashed trace) field sweeps is seen at $\sigma =1$.
}
\label{kumadafig1}
\end{center}
\end{figure}

In Fig.\,\ref{kumadafig1}, we show how the $\nu =2/3$ state evolves as the total electron density $n_t$ changes in the perpendicular field ($B_{tot}=B_\perp$), where $B_{tot}$ is the total magnetic field. 
Data are given for two limiting cases, i.e. the balanced density point ($\sigma =0$) and the monolayer point ($\sigma =1$).
At both points, the magnetoresistance $R_{xx}$ becomes zero at higher density, where QH states are well developed.
At $\sigma =0$, the minimum of $R_{xx}$ gets weaker with decreasing $n_t$, and collapses at $n_t=0.8\times 10^{11}$\,cm$^{-2}$, followed by a reappearance of the $\nu =2/3$ minimum at $n_t=0.6\times 10^{11}$\,cm$^{-2}$ \cite{Suen,Lay}.
At $\sigma =1$, on the other hand, we observed a totally different behavior: A significant difference in $R_{xx}$ between upward (solid trace) and downward (dashed trace) magnetic-field sweeps was observed for a wide range of magnetic field covering the entire $\nu =2/3$ region.
We call it type-G hysteresis (G stands for the ground state, as we explain later).

\begin{figure}[t]
\begin{center}
\includegraphics[width=60mm]{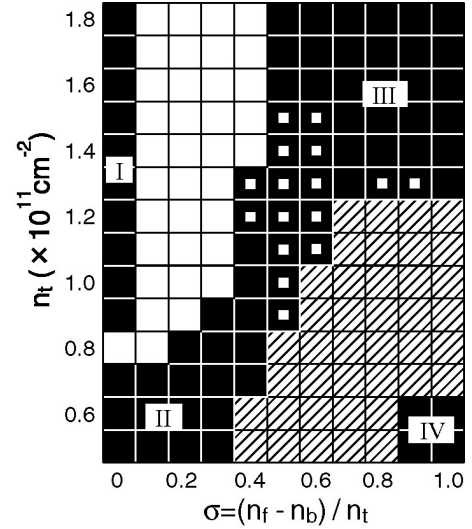}
\caption{\small
Phase diagram illustrated for $\nu =2/3$.
The horizontal axis is the normalized density difference between two layers $\sigma $ and the vertical axis is the total electron density $n_t$.
$R_{xx}$ minimum for QH state was found in the black area.
QH effect was not observed in the white area.
Type-G hysteresis was observed in the hatched area as in Fig.\,\ref{kumadafig1}.
The small white square in black region indicates occurrence of different type of hysteresis, where a QH state is developed and the hysteresis is found at fields slightly lower than that at $\nu =2/3$ (Fig.\ref{kumadafig5}).}
\label{kumadafig2}
\end{center}
\end{figure}

We repeated similar measurements for $0\leq \sigma \leq 1$ in the range of $0.5\leq n_t\leq 1.8\times 10^{11}$\,cm$^{-2}$ and made a phase diagram (Fig.\,\ref{kumadafig2}) \cite{QH}.
In the phase diagram, four QH (black) areas labeled I, \mbox{\RB}, \mbox{\RC} and \mbox{\RD} are clearly recognized.
Area I is completely separated by the no-QH (white) area, while area \mbox{\RD} is connected with areas \mbox{\RB} and \mbox{\RC} by the type-G hysteresis (hatched) region.
On the other hand, the low density phase at $\sigma =0$ (area \mbox{\RB}) evolves continuously into the high density phase at $\sigma =1$ (area \mbox{\RC}).
In these two areas (\mbox{\RB} and \mbox{\RC}), we observed a new type of hysteresis (small white squares in the black) region, which we name type-E hysteresis (E stands for excitation levels).
We discuss this hysteresis later in Fig.\,\ref{kumadafig5}.

\begin{figure}[b]
\begin{center}
\includegraphics[width=70mm]{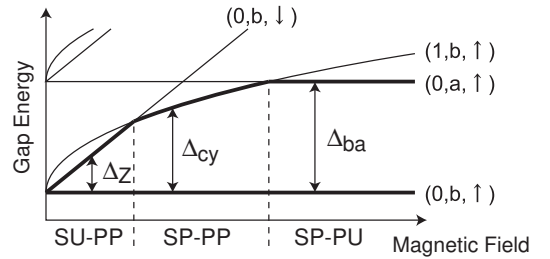}
\caption{\small
Low-lying energy levels of CF are depicted schematically as a function of the magnetic field.
At $\nu =2/3$, the lowest two levels are filled up (thick curves), and three QH states are expected to appear.
}
\label{kumadafig3}
\end{center}
\end{figure}

We explain these four QH areas based on the CF model extended to a bilayer system.
CF energy levels are split by the Zeeman $\Delta _Z$, pseudo-Zeeman $\Delta _{ba}$ and CF-cyclotron $\Delta _{cy}$ energies (Fig.\,\ref{kumadafig3}).
We use the pseudospin language to deal with the layer degree of freedom.
The pseudo-Zeeman energy is the energy gap between the bonding (b) and antibonding (a) states \cite{MurakiSSC}, $\Delta _{ba}=\Delta _{SAS} \sqrt{1/(1-\sigma ^2)}$, which is equal to $\Delta _{SAS}$ at $\sigma =0$ and increases with applying a bias voltage.
The origin of the CF-cyclotron gap is a Coulomb interaction \cite{HLR}, and we set $\Delta _{cy}=C(\sigma )e^2/4\pi \epsilon l_B^\ast \propto \sqrt{B_\perp }$, where $C(\sigma )$ is a dimensionless coefficient, $\epsilon $ is the dielectric constant, $l_B^\ast $ is the magnetic length in the effective magnetic field $B^\ast =|B_\perp -B_{1/2}|$ and $B_{1/2}$ is the magnetic field at $\nu =1/2$.
We label each CF energy level as ($N_{\rm CF},ps,s$), where $N_{\rm CF}$ $(=0,1)$, $ps$ $(={\rm b},{\rm a})$ and $s$ $(=\uparrow ,\downarrow)$ are the CF-Landau orbit, pseudospin and spin indices.
At $\nu =2/3$, two CF energy levels are occupied.

When $n_t$ is high at the balanced density point (area I), 
$\Delta _Z$ is large and $\Delta _{ba}$ is small, 
and then the spin-polarized and pseudospin-unpolarized (SP-PU) state is realized.
In this state, the intralayer interaction dominates the interlayer interaction, and in each layer electrons configure the monolayer $\nu =1/3$ QH state.
When the density is off-balanced ($\sigma \neq 0$), the filling factor in the front and back layers deviate from 1/3, and the compound state collapses.
Thus, this state is stable only at $\sigma =0$.
In area \mbox{\RD}, where $n_t$ is small and $\sigma $ is large, the spin-unpolarized and pseudospin-polarized (SU-PP) state is realized.
When $n_t$ is increased from area \mbox{\RD} to \mbox{\RC}, the SU-PP state transits to the spin-polarized and pseudospin-polarized (SP-PP) state.
Even at $\sigma =0$, $\Delta _{ba}$ has a finite value $\Delta _{SAS}$, which exceeds the CF-cyclotron gap at low density.
Thus, the QH state in area \mbox{\RB} is the SP-PP state.
Accordingly, the SP-PP state persists continuously from area \mbox{\RB} to \mbox{\RC}.

\begin{figure}[t]
\begin{center}
\includegraphics[width=80mm]{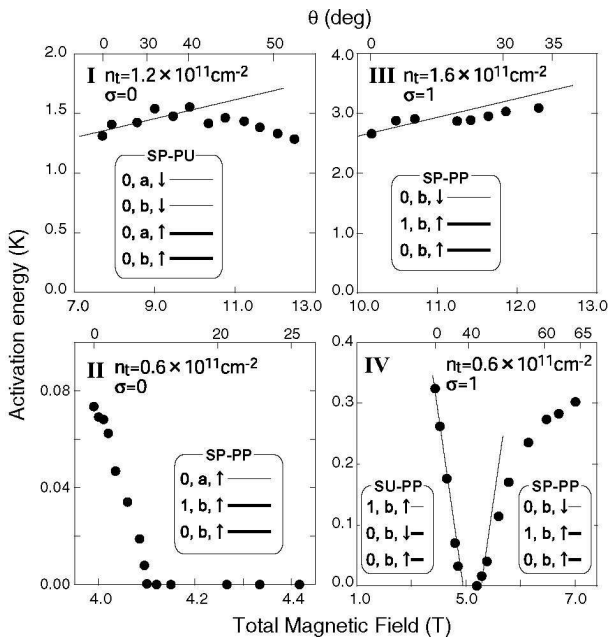}
\caption{\small
The activation energy by tilting the sample as a function of $B_{tot}$ at $\nu =2/3$ in the four QH areas.
On the top axis, we show the tilt angle $\theta $.
The leftmost point is for the data at $\theta =0$ where $B_{tot}=B_\perp $.
The slope of the lines included in I, \mbox{\RC} and \mbox{\RD} correspond to $\pm g^\ast \mu _BB_{tot}$.
In the insets, we showed the ground state of the QH states (thick line) and excitation levels (thin line).}
\label{kumadafig4}
\end{center}
\end{figure}

To elucidate these bilayer QH states more in detail, we measured the activation energy $\Delta $ by tilting the sample in the magnetic field with keeping $B_\perp $ fixed, where $\Delta $ is determined from the temperature dependence of the magnetoresistance \cite{LK}: $R_{xx}\propto \exp (-\Delta /2T)$.
Figure\,\ref{kumadafig4} shows the activation energy in the four QH areas as a function of the total magnetic field $B_{tot}=(B_\perp ^2+B_\parallel ^2)^{1/2}$, where $B_\parallel$ is the in-plane magnetic field.
As $B_\parallel $ is increased, $\Delta _Z=g^\ast \mu _BB_{tot}$ increases and $\Delta _{SAS}\propto \exp\{-(B_\parallel d/B_\perp l_B^\ast )^2\}$ decreases \cite{Hu}, where $g^\ast $ is the gyromagnetic ratio, $\mu _B$ is the Bohr magneton and $d$ is the layer separation.
In area I and \mbox{\RC}, $\Delta $ initially increases, which indicates that the excitation gap is the Zeeman energy.
In area \mbox{\RB}, when $B_{tot}$ is increased only by 0.1\,T, the activation energy becomes zero.
Although we increased $B_{tot}$ up to 13\,T, a QH state did not reappear.
The excitation gap of this state is $\Delta _{ba}$, because $B_\parallel $ decreases $\Delta _{SAS}$.
In area \mbox{\RD}, the activation energy first decreases and shows a pronounced transition at 5\,T.
This behavior is interpreted as the phase transition from the SU-PP to SP-PP states due to the increased $B_\parallel $ \cite{Eisenstein,Engel}.
This SP-PP state is the same as the one in area \mbox{\RC}.

We analyze the phase transition points based on the noninteracting CF model (Fig.3).
We start with the phase transition point between the SP-PP (area \mbox{\RB} and \mbox{\RC}) and SP-PU (area I) states at the balanced point ($\sigma=0$).
It occurs due to the crossing of the levels (1,b,$\uparrow $) and (0,a,$\uparrow $) at
\begin{equation}
\Delta _{ba}=C(\sigma )e^2/4\pi \epsilon l_B^\ast.
\label{csas}
\end{equation}
The transition occurs at $n_t=0.8\times 10^{11}$\,cm$^{-2}$ in our data (Fig.\,\ref{kumadafig2}).
As mentioned above, the SP-PU state made of (0,b,$\uparrow$) and (0,a,$\uparrow$) does not exist in the unbalanced configuration ($\sigma\not=0$) and the no-QH area develops between the areas I and \mbox{\RC} as in the white area in Fig.\,\ref{kumadafig2}.
Then, eq.\,(\ref{csas}) determines the phase boundary between the SP-PP state and the no-QH area. 
The transition point shifts to higher $n_t$ as $\sigma $ is increased, $n_t \propto 1/\big(C^2(\sigma)(1-\sigma^2)\big)$.

Next, we discuss the transition point between the SP-PP (area \mbox{\RB} and \mbox{\RC}) and SU-PP (area \mbox{\RD}) states, where the Zeeman $\Delta _Z$ and CF-cyclotron $\Delta _{cy}$ gaps become equal,
\begin{equation}
g^\ast \mu _BB_\perp =C(\sigma )e^2/4\pi \epsilon l_B^\ast .
\label{czee}
\end{equation}
The transition occurs at $n_t=0.95\times 10^{11}$\,cm$^{-2}$ in the monolayer limit ($\sigma =1$) in our data (Fig.\,\ref{kumadafig2}). 
When $\sigma $ decreases, $C(\sigma )$ decreases because electrons tend to extend over both of the layers and the Coulomb energy decreases.
It follows from eq.\,(\ref{czee}) that $B_{\perp }\propto C(\sigma)^2$.
Hence, the total density at the transition point decreases as $n_t=\nu eB_\perp /h\propto C(\sigma )^2$ \cite{SUPP}.
This explains why the SP-PP state is realized at higher density near the monolayer point (area \mbox{\RC}) but at lower density around the balanced point (area \mbox{\RB}).

In this way, the phase diagram for $\nu =2/3$ is interpreted qualitatively 
by the noninteracting CF model.
However, some refinement is needed quantitatively.
We focus on the coefficient $C(\sigma )$ of the CF cyclotron gap.
From eq.\,(\ref{czee}) we obtain $C(1)=0.026$ at $\sigma =1$.
We have argued that it decreases as $\sigma$ decreases,
and hence $C(0)$ must be much smaller than $0.026$,
as is suggested in our phase diagram (Fig.\,\ref{kumadafig2}) \cite{SUPP}.
However, from eq.\,(\ref{csas}) we obtain $C(0)=0.17$ at $\sigma =0$,
which is almost ten times bigger than the one implied by eq.(1).
We have reached this inconsistency by assuming that 
collective Coulomb interactions renormalize only the CF cyclotron energy 
as in the noninteracting CF theory of the monolayer FQHE. 
This simple picture fails in the bilayer FQHE.
The above inconsistency would be resolved by the understanding that collective Coulomb interactions renormalize also the Zeeman and tunnelling energies of CFs.

\begin{figure}[t]
\begin{center}
\includegraphics[width=60mm]{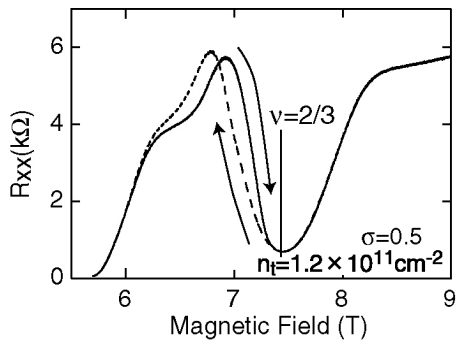}
\caption{\small
The magnetoresistance around $\nu =2/3$ at $n_t=1.2\times 10^{11}$\,cm$^{-2}$ at $\sigma =0.5$.
The $\nu =2/3$ QH state is well developed and type-E hysteresis is seen only at the fields slightly lower than that at $\nu =2/3$.}
\label{kumadafig5}
\end{center}
\end{figure}

Finally, we discuss the two types of the hysteresis.
Type-G hysteresis was observed between the SU-PP and SP-PP states.
This type of hysteresis has already been observed in various monolayer systems, and argued to be caused by domain morphology in the ground state \cite{Cho,Eom,Smet}.
The ground state consists of the spin-polarized and spin-unpolarized domains at the crossover point of the two states.
The hysteresis would result from a conspiracy between electronic and nuclear polarization \cite{Kronmuller1,Kronmuller2,Hashimoto}.

Type-E hysteresis is distinguished from type-G one by the fact that the ground state does not change. 
Thus, type-E hysteresis is not due to the domain morphology of the ground state.
We notice that type-E hysteresis appears not exactly at $\nu=2/3$ but only on the lower field side of $\nu =2/3$ (Fig.\,\ref{kumadafig5}).
We speculate the origin of the hysteresis as follows.
At the lower field side of $\nu =2/3$, many quasiparticles are created in excitation levels of $\nu =2/3$.
The SP-PP state in areas \mbox{\RB} and \mbox{\RC} have different excitation levels (0,a,$\uparrow $) and (0,b,$\downarrow $), as shown in the insets of Fig.\,\ref{kumadafig4}.
These upper CF levels are almost degenerate with one another in the small white square area in the phase diagram (Fig.\,\ref{kumadafig2}).
We would conclude that type-E hysteresis occurs due to the domain morphology involving different spin excitation levels.
As far as we are aware of, type-E hysteresis has not been observed before. 

In summery, we made the phase diagram for $\nu =2/3$ by changing the total electron density and the density difference between two layers.
In the phase diagram, three different spin/pseudospin QH states were observed.
The SP-PP state is realized at lower density around the balanced point, while at higher density near the monolayer point.
A no-QH area develops between the SP-PU and SP-PP states.
Moreover, two types of hysteresis were observed.
One is due to the crossover of different spin ground states, and the other is associated with crossing of different spin excitation levels.
We have interpreted the phase diagram by a CF model extended to a bilayer system.
These phases are caused by the spin, layer and composite fermion Landau orbital degrees of freedom.

The research was supported in part by Grant-in-Aids for the Scientific Research from the Ministry of Education, Science, Sports and Culture (Nos. 10203201, 11304019, 08159), the Mitsubishi Foundation and the Asahi glass Foundation, CREST-JST and NEDO "NTDP-98" projects.

\scriptsize


\begin{thebibliography}{99}
\bibitem{cyc}
In GaAs, the cyclotron energy is about 60 times lager than the Zeeman energy when $B_\perp =B_{tot}$.
\bibitem{QH}
Though the hysteresis depends on the magnetic sweep rate and the current, we selected suitable condition (0.06\,T/min, 20\,nA) to judge whether the hysteresis exists.
\bibitem{SUPP}
Note that, though the QH state in the area \mbox{\RD} has been observed only at a low $n_t$ and large $\sigma $ corner in Fig.\,\ref{kumadafig2}, it is expected to extend toward $\sigma =0$ for lower $n_t$: QH states are hard to observe at lower density, $n_t\leq 0.4\times 10^{11}$\,cm$^{-2}$, where the mobility is lowered.
\end{thebibliography}
\end{document}